\documentclass{ccjnl}

\title{Passive Integrated Sensing and Communication Scheme based on RF Fingerprint Information Extraction for Cell-Free RAN}
\author{Jingxuan Yu\inst{1}, Fan Zeng\inst{1}, Jiamin Li\inst{1,2,*}, Feiyang Liu\inst{3}, Pengcheng Zhu\inst{1,2}, Dongming Wang\inst{1,2}, Xiaohu You\inst{1,2}\corinfo{jiaminli@seu.edu.cn}}
\receiveddate{}
\reviseddate{}
\Editor{}

\address[1]{National Mobile Communications Research Laboratory, School of Information Science and Engineering, Southeast University, Nanjing 210096, China}
\address[2]{Purple Mountain Laboratories, Nanjing 211111, China}
\address[3]{54th Research Institute of Electronics Technology Group Corporation, Shijiazhuang 050081, China}

\begin{document}
\maketitle

\begin{abstract}
This paper investigates how to achieve integrated sensing and communication (ISAC) based on a cell-free radio access network (CF-RAN) 
architecture with a minimum footprint of communication resources. We propose a new passive sensing scheme. The scheme is based on the radio frequency (RF) 
fingerprint learning of the RF radio unit (RRU) to build an RF fingerprint library of RRUs. The source RRU is identified by comparing 
the RF fingerprints carried by the signal at the receiver side. The receiver extracts the channel parameters from the signal and estimates 
the channel environment, thus locating the reflectors in the environment. The proposed scheme can effectively solve the problem of interference 
between signals in the same time-frequency domain but in different spatial domains when multiple RRUs jointly serve users in CF-RAN architecture. 
Simulation results show that the proposed passive ISAC scheme can effectively detect reflector location information in the environment without 
degrading the communication performance. 
\end{abstract}

\section{introduction}
\label{s1}

With the advancement of time and technology, mobile communication and radio sensing technologies are flourishing, the 
similarities between the radio frequency (RF) architectures used in communication and radar are increasing, and the frequency bands used are gradually 
approaching or even overlapping \cite{1}. There is a growing interest in the field of information research in integrated sensing and 
communication (ISAC). The two main types of ISAC are active and passive sensing \cite{2}, where active sensing means that the object being 
sensed has RF communication capabilities, while passive sensing works more like radar \cite{3,4}. In previous studies, sensing 
signals were specifically designed or only single base station sensing was considered, which takes up possible communication channels 
and has an impact on communication rates in future mobile communications \cite{5,6,7,8}. 

In the next generation of mobile communications research, researchers have proposed a new communication service architecture, the cell-free 
radio access network (CF-RAN), which eliminates inter-cell signal interference through innovations to the cell structure \cite{9,10}. CF-RAN is 
a system of a large number of RF radio units (RRUs) distributed over a large area, serving a large number of user terminals at the same time or frequency 
band, and is able to improve the service upstream and downstream rates of users. Passive sensing techniques based on multi-input multi-output (MIMO) antenna RF devices 
require that the relative positions of the transmitter and receiver are known and that the signals used for sensing are known to the 
receiver \cite{6}. The CF-RAN architecture provides exactly all the basic conditions needed for passive sensing. In this paper, a CF-RAN 
architecture in network assisted full duplex (NAFD) mode is used, where the RRUs within the cluster are designated as uplink or downlink 
respectively at any given time slot, and the console can rationally deploy uplink and downlink resources, resulting in better communication 
and sensing performance \cite{11,12}. 

Since a large number of RRUs transmit and receive signals with the same content, distinguishing different signal sources at the receiver 
side becomes a key point to solve passive sensing in the CF-RAN architecture. In CF-RAN architecture, multiple RRUs use the same waveform 
containing communication data for passive sensing at the same time, and the receiver receives the same content of communication waveform 
from multiple RRUs. It is necessary to distinguish the transmitting source of each signal in order to determine the relative position of 
transmitter and receiver, otherwise, effective passive sensing cannot be performed. RF fingerprinting is a good tool to effectively 
distinguish different sources without taking up RF resources. 

RF fingerprinting is a process of identifying RF devices or judging unfamiliar RF devices. Due to the specificity of the wireless 
transmitter circuit and the imperfection of the components that make up the transmitter circuit, it is possible to extract the RF 
fingerprint used from the wireless signal to identify the wireless transmitter \cite{13}. The RF fingerprint is the physical layer nature of the wireless 
communication device. It is a characteristic that is included in every signal sent by an RF device after it has been manufactured. 
It was shown that this feature can be used to distinguish and identify RF hardware from the same manufacturer with the same 
model number and different serial numbers \cite{14}. For communication signals, features can be extracted directly from their time-domain 
waveforms, and it is also possible to process the communication signal in the transform domain and then extract the features. Intelligent 
algorithms based on deep convolutional neural networks (CNN) \cite{13} and hybrid neural networks \cite{15} have also been proposed for RF fingerprint extraction of signals.   

This paper investigates the passive integrated sensing and communication for cell-free RAN. The main contributions of this paper are summarized as follows:
\begin{enumerate}

  \item{A multi-RRU joint ISAC architecture is proposed in the CF-RAN framework with network-assisted full-duplex mode. The working mechanism of passive sensing in this architecture is discussed, and the issue of RF source identification in simultaneous same-frequency passive sensing is analyzed.}
  \item{The interference problem of co-channel interference in multi-RRU joint ISAC service is discussed. Effective passive sensing is achieved by utilizing the characteristics of RF fingerprint identification, which does not occupy the communication bandwidth and does not degrade the communication rate and latency. A positioning scheme for passive reflectors is modeled and verified through simulation.}
  \item{The feasibility of RF fingerprint identification in distinguishing signals from multiple RRUs is analyzed in this architecture. The effectiveness of RF fingerprint identification is simulated and validated using a CNN (Convolutional Neural Network) model, and the relationship between passive sensing accuracy and estimation errors of different levels of sensing parameters is obtained. Simulation results compare the performance parameters of ISAC in NAFD mode and single downlink base station mode.}
  
\end{enumerate}

\section{System model}
\subsection{System description}
\label{s2}

\begin{figure}[!ht]
\centering
\includegraphics[width=1\linewidth]{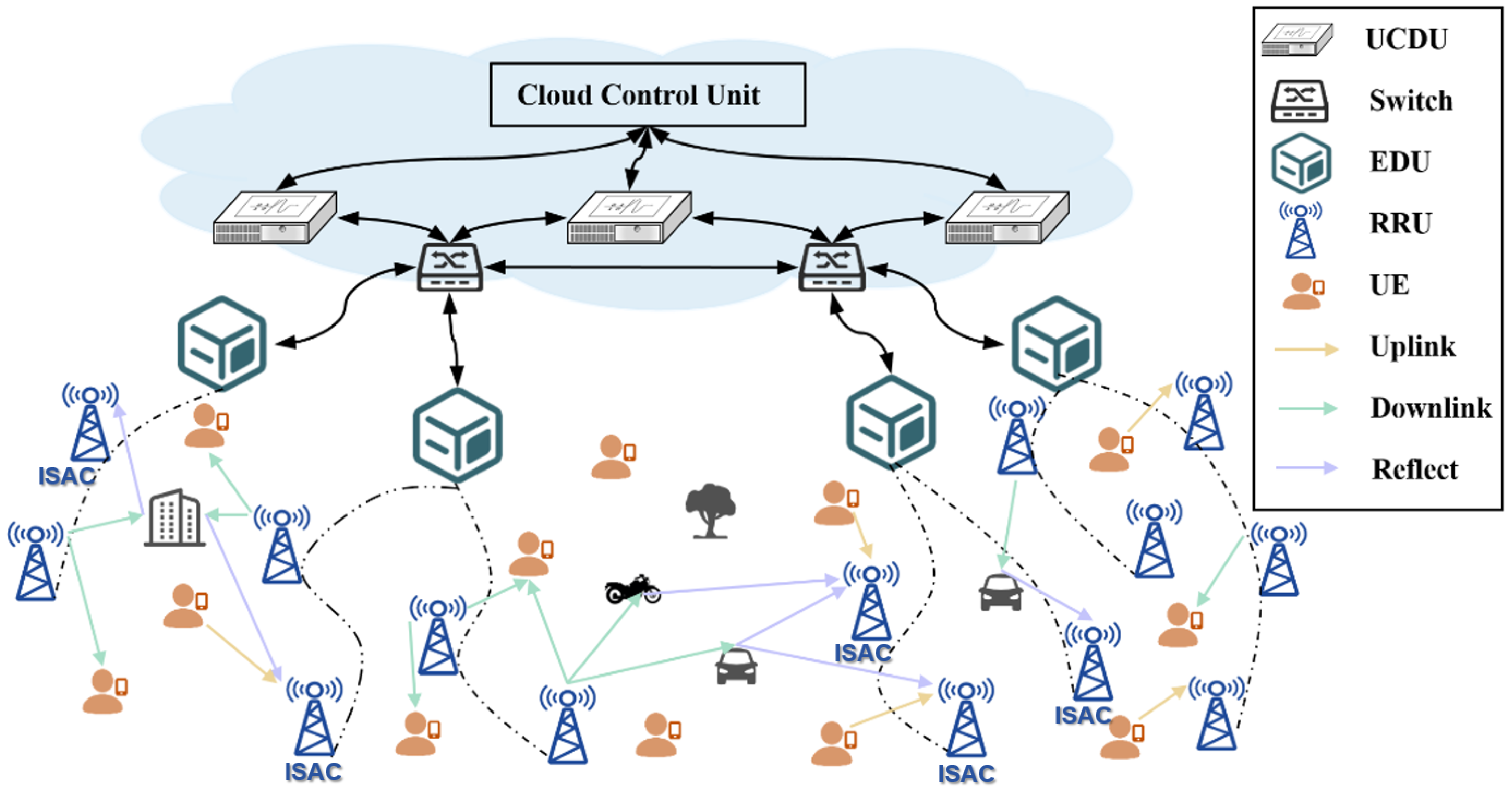}
\caption{CF-RAN NAFD ISAC architecture diagram.}
\label{fig1}
\end{figure}

As in Figure~\ref{fig1}, considering a CF-RAN ISAC architecture, in a given area, multiple spatially adjacent RRUs are connected to a common edge distributed unit (EDU). The EDU is responsible for the RF signal processing and computation tasks of the associated RRUs, as well as the allocation of the uplink and downlink operation modes of the RRUs. The EDU can determine which RRUs have ISAC functionality, while others focus solely on communication services, thus striking a balance between perception and communication performance. The EDU is connected to the user-centric distributed unit (UCDU) in the cloud through a switch, which possesses significant computational resources for handling complex tasks and overall resource scheduling capabilities. Each RRU is equipped with a uniform MIMO antenna array of size $M \times N$, while the user equipment (UE) has a single antenna. At a given time, $D$ downlink RRUs transmit downlink ISAC signals.
The uplink RRU receives the uplink signal $\boldsymbol{y}_1$ sent by the UE terminal and the downlink signal $\boldsymbol{y}_2$ from the downlink RRU. Due to the presence of reflectors, $\boldsymbol{y}_2$ contains multiple non line of sight (NLOS) signals. Reflectors can include low-flying aircraft, vehicles traveling on the road, as well as buildings and long-standing facilities that serve as fixed reflectors. These fixed reflectors can be easily filtered and eliminated by the RRU through observation over a period of time. Once an RRU obtains perception information, it can share the location information with other EDUs and RRUs, thereby improving overall operational efficiency and optimizing decision-making.
    
\subsection{Channel and data model}
Considering the uplink communication process between the RRUs and terminals under the control of an EDU, within a time period, the EDU allocates $U$ uplink RRUs and $D$ downlink RRUs. Under normal operating conditions, the channels related to the $u$-th uplink RRU consist of the uplink channel from the UE to the uplink RRU and the downlink channel from the downlink RRU to the uplink RRU.

The channel $\mathbf{H}_{d, u}$ between the $d$-th downlink RRU and the $u$-th uplink RRU can be represented as follows:

\begin{equation}
  \label{eq1}
  \mathbf{H}_{d, u}=\sum_{l=1}^{L} \mathbf{h}_{d, u, l},
  \end{equation}
    
  \begin{equation}
      \label{eq2}
      \mathbf{h}_{d, u, l}=\alpha_{d, u, l} \boldsymbol{e}^{-j 2 \pi {f}_{0} \tau_{d, u, l} } \boldsymbol{e}^{j 2 \pi T_{x} \nu_{d, u, l} } \boldsymbol{a}\left(\varphi_{d, u, l}, \theta_{d, u, l}\right).
  \end{equation}
      
Eq.~\eqref{eq1} contains the line of sight (LOS) path channel and the NLOS path channel, NLOS path is generated by the reflection of a reflector. 
$\boldsymbol{e}^{-j 2 \pi f_{0} \tau_{d, u, l} }$  represents the phase shift caused by delay,$\tau_{d, u, l}$represents the propagation delay, $f_{0}$ represents the frequency of the signal, $\boldsymbol{e}^{j 2 \pi T_{x} \nu_{d, u, l} }$ is Doppler frequency shift, $T_{x}$ is sampling frequency, $\nu_{d, u, l}$ is the Doppler parameter, $\alpha_{d, u, l}$ is the fading coefficient, and $\boldsymbol{a}\left(\varphi_{d, u, l}, \theta_{d, u, l}\right)$ 
is the vector coefficient of the channel, which includes the information of the channel's transmission, reflection, and incident angles. Similarly, the channel $\mathbf{H}_{q, u}$ between the $q$-th uplink UE and the $u$-th uplink RRU can be represented similarly to Eq.~\eqref{eq1}~\eqref{eq2}.

For transmitting signals, the downlink data is denoted as $\boldsymbol{s}=\left[s_{1}, \ldots, s_{N_{A}}\right]^{\mathrm{T}}$. $\boldsymbol s$ 
contains $N_{A}$ data symbols. The signal emitted by the $d$th downlink RRU can be expressed as \cite{14}

\begin{equation}
\label{eq3}
\boldsymbol{x}_{d}=\boldsymbol{F}_{d} \boldsymbol{\gamma}_{d}^{\frac{1}{2}} \boldsymbol{s} \boldsymbol{\gamma}_{d}^{\frac{1}{2}} \boldsymbol{f}_{d},
\end{equation}
where $\boldsymbol{F}_{d}$ is the beamforming matrix, $\boldsymbol{\gamma}_{d}^{\frac{1}{2}}$ is the diagonal array containing the power 
control coefficients, and $\boldsymbol{f}_{d}$ is the combined value of multiple nonlinear integrals present in the signal, 

which represent the fingerprint characteristics introduced by RF components in the RRU. The origin of these fingerprint characteristics lies in the hardware imperfections of the transmitting devices, including clock jitter, DAC sampling errors, nonlinearity of mixers or local frequency synthesizers, nonlinearity of power amplifiers, characteristics of equipment antennas, and sub-circuits of analog modulators. These hardware imperfections contribute to the generation of multiple fingerprint features in the RF signal. It is worth noting that the same transmitter exhibits similar RF fingerprint noise for signals with different loads, indicating a consistent pattern in the RF fingerprint characteristics \cite{wang2016wireless}.

For the uplink RRU receiving the signal from the uplink UE and the signal from the downlink RRU, the signal received at the $u$-th uplink RRU 
can be expressed as

\begin{equation}
\label{eq4}
  \begin{array}{l}
    \boldsymbol{y}_{u}=\sum_{q=1}^{Q} \mathbf{H}_{q, u} \boldsymbol{x}_{q, u}+\sum_{d=1}^{D} \mathbf{H}_{d, u} \boldsymbol{x}_{d, u}+\mathbf{w}_{u} \\
    =\underbrace{\sum_{q=1}^{Q} \mathbf{H}_{q, u} \boldsymbol{x}_{q, u}}_{X_{UE}}+\underbrace{\sum_{d=1}^{D} \mathbf{h}_{d, u, 1} \boldsymbol{x}_{d, u}}_{Y_{LOS}}+\underbrace{\sum_{d=1}^{D} \sum_{l=2}^{L} \mathbf{h}_{d, u, l} \boldsymbol{x}_{d, u}}_{\hat{Y}_{NLOS}+\check{Y}_{NLOS}}+\mathbf{w}_{u}, 
  \end{array}
\end{equation}

Where the channel between the $q$-th UE and the $u$-th uplink RRU is denoted as $\mathbf{H}_{q, u}$, $\boldsymbol{x}_{q, u}$ is the uplink data of the $q$-th user, $\mathbf{H}_{d, u}$ is the channel between the $d$-th downlink RRU and the $u$-th uplink RRU, and $\boldsymbol{x}_{d, u}$ is the data signal containing RF fingerprint information from the $d$-th RRU. 
$\mathbf{h}_{d, u, 1}$ represents the LOS path channel. $\mathbf{h}_{d, u, l}$ represents the NLOS channels due to passive object reflections, where there are $L-1$ such channels.

Among them, $\hat{Y}_{NLOS}$ represents the NLOS path caused by long-standing objects such as the ground or buildings, while $\check{Y}_{NLOS}$ represents mobile reflectors such as vehicles. $\mathbf{w}_{u}$ represents Gaussian noise. 

Due to the nature of the CF-RAN architecture, the channel $\mathbf{h}_{d, u, 1}$ and data $\boldsymbol{x}_{d, u}$ are known for the RRU, while the unknown information consists of the uplink user channel $\mathbf{H}_{q, u}$, data $\boldsymbol{x}_{q, u}$, and channel $\mathbf{h}_{d, u, l}$.
The received uplink signal at the EDU is represented as $\boldsymbol{y}=\left[\boldsymbol{y}_{1},\boldsymbol{y}_{2}, \ldots, \boldsymbol{y}_{U}\right]$.

To avoid interference from the uplink UE channels $\mathbf{H}_{q, u}$,  the EDU performs ISAC passive sensing during the uplink idle state. The uplink idle state is defined as the absence of uplink UE signals that can interfere with the downlink ISAC signals used for passive sensing within the same frequency band.
The EDU determines whether it is in the uplink idle state based on a power threshold.
For the received uplink signal $\boldsymbol{y}$ at the EDU, we subtract the known downlink signals $\hat{Y}_{NLOS}$ and $\check{Y}_{NLOS}$ transmitted by the downlink base stations to obtain $\hat{\boldsymbol{y}}_{u} = \boldsymbol{y}_{u} - Y_{LOS} - \hat{Y}_{NLOS} = X_{UE} + \check{Y}_{NLOS} + \mathbf{w}_{u}$. The remaining NLOS path signal power, which experiences higher energy loss due to reflections, can be evaluated using a power threshold to determine if there is any signal $X_{UE}$ being received.

When the uplink is idle, the $u$-th RRU's received signal $\boldsymbol{y}_{u}=\underbrace{\sum_{d=1}^{D} \mathbf{h}_{d, u, 1} \boldsymbol{x}_{d, u}}_{Y_{LOS}}+\underbrace{\sum_{d=1}^{D} \sum_{l=2}^{L} \mathbf{h}_{d, u, l} \boldsymbol{x}_{d, u}}_{Y_{NLOS}}+\mathbf{w}_{u}$ is a portion of Eq.~\eqref{eq4} . 
When processing the received signal, we only need to consider the identical downlink signals from multiple transmitters, like $\hat{\boldsymbol{y}}_{u} = \boldsymbol{y}_{u} - Y_{LOS} - \hat{Y}_{NLOS} = \check{Y}_{NLOS} + \mathbf{w}_{u}$.
We take the LOS path signal $Y_{LOS}$ from the downlink RRU as the reference signal. In addition to the reference signal, the signal $\boldsymbol{y}_{u}$ also includes the downlink signals $Y_{NLOS}$ reflected by multiple reflectors due to $D$ transmitters. Among them, $\hat{Y}_{NLOS}$ represents the signals caused by long-standing objects, which can be learned and eliminated. The remaining signal $\check{Y}_{NLOS}$ represents the passive reflectors in the environment that need to be sensed. Previous research has explored various methods, such as tensor decomposition \cite{gong2022multipath,lin2021tensor,zanatta2020tensor}, to differentiate the multipath information to obtain the arrival angles, Doppler information, and time delays relative to the reference signal. With this information, the EDU can estimate the distance of signal transmission and obtain the relative distance of the detected passive reflectors.

\section{ISAC solution based on RF fingerprint extraction}
\subsection{Separation of multipath information}

Since we specify that the EDU performs ISAC sensing during uplink idle periods, we can primarily focus on the analysis of the downlink RRU signals. As discussed in the previous chapter, the signals from the downlink RRUs form NLOS paths through reflections in the environment, providing perception information to the receivers. Therefore, for passive sensing, we consider the signals received by each individual downlink RRU. The joint processing of sensing signals is possible, and the computation is performed at the EDU. For downlink passive sensing, each RRU receives multipath signals from the other $D$ downlink RRUs. For the $u$-th uplink RRU, the received signal can be represented as \cite{19}

  \begin{align}
    \label{eq6}
    \boldsymbol{y}_{u}=&\sum_{d=1}^{D} \mathbf{h}_{d, u} \boldsymbol{x}_{d} + \mathbf{w}_{u} \nonumber \\
    &=\sum_{d=1}^{D} \sum_{l=1}^{L} \alpha_{d, u, l} \boldsymbol{e}^{-j 2 \pi f_{0} \tau_{d, u, l}} \boldsymbol{e}^{j 2 \pi T_{x} \nu_{d, u, l} } \nonumber \\
    &\boldsymbol{a}\left(\varphi_{d, u, l}, \theta_{d, u, l}\right) \boldsymbol{x}_{d}+\mathbf{w}_{u} \nonumber \\
    &=\underbrace{\mathbf{G} \mathbf{A}(\phi, \theta) \mathbf{D} \mathbf{C}}_{\mathbf{H}_{u}} \boldsymbol{x}_{d}+\mathbf{w}_{u},
  \end{align}

The subscript $d$ represents the $d$-th RRU, $\boldsymbol{x}_{d}$ is the signal transmitted by the $d$-th RRU on the subcarrier. 
$\mathbf{A}(\phi, \theta)$ is a block-diagonal matrix of size $MND \times L$, the diagonal elements of $\mathbf{D}$ and $\mathbf{C}$ are $\boldsymbol{e}^{-j 2 \pi f_{0} \tau_{d, u, l}}$ and $\boldsymbol{e}^{j 2 \pi T_{x} \nu_{d, u, l} }$, $\mathbf{w}_{u}$ is the noise vector. The term fingerprint in Eq.~\eqref{eq6} represents the transmitted data containing RF fingerprint information.

The model in Eq.~\eqref{eq6} has a similar channel structure representation as the basic model in Eq.~\eqref{eq2}, but the specified multipath signal is directed 
to a different RRU. The paper \cite{18} verifies that there is always a sufficient number of antennas to satisfy the performance requirements in a massive MIMO 
radar system, without any a priori perturbation statistics. For the method of distinguishing multipath information, the five sensing parameters 
in Eq.~\eqref{eq2} can be estimated separately, or jointly by building 1-D to 4-D CS models \cite{19}. The estimation of sensing parameters is not the focus of 
this paper and will not be discussed in depth here. But it is a point worthy of continued research and perhaps dedicated to more optimal solutions in our 
future studies.

From equation Eq.~\eqref{eq6}, it can be seen that encapsulating $\boldsymbol{y}_{u}$ from multiple RRUs increases its 
length, but the unknown parameters increase similarly. Therefore, sensing does not directly 
benefit from the joint processing of the received signal as a whole. However, the parameters 
of the signal propagation between RRUs may be similar due to channel reciprocity. Such a 
property can be used for joint processing across RRUs. The prerequisite for joint processing is 
the identification of the RF fingerprint information contained in the $\mathbf{x}_{d}$. 

\subsection{RF fingerprint extraction}
In the previous discussion we have already mentioned possible solutions for passive sensing. 
Method is extracting the RF fingerprint information from the cell-free ISAC RRU, 
identifying the RF fingerprint to determine the source of the signal. 
The objectives to be focused on in this section are the training and extraction of fingerprint 
information, the parsing of the receiver signal and the localisation of passive objects.

In Section 2.2, we modeled the RF fingerprint information of the transmitted signal and discussed how $\boldsymbol{f}_d$ is generated as a combination of multiple nonlinear integrals in the signal. Numerous studies have been conducted based on the fingerprint characteristics introduced by one or more RF components, mapping these fingerprint features to the RF components themselves using traditional or intelligent design methods. The goal is to match the received signal with the known transmitter based on these inherent hardware features, aiming to determine if it is from a familiar transmitter. This technology has produced many useful results and conclusions.
Based on existing RF fingerprint extraction techniques, we have designed a classifier that utilizes tensor decomposition algorithms to separate the received multipath signals. By inputting the separated multipath signals into the classifier, the classifier outputs the ID of the transmitting RRU, thus identifying the source of the received signal.

\begin{figure}[!ht]
\centering
\includegraphics[width=0.9\linewidth]{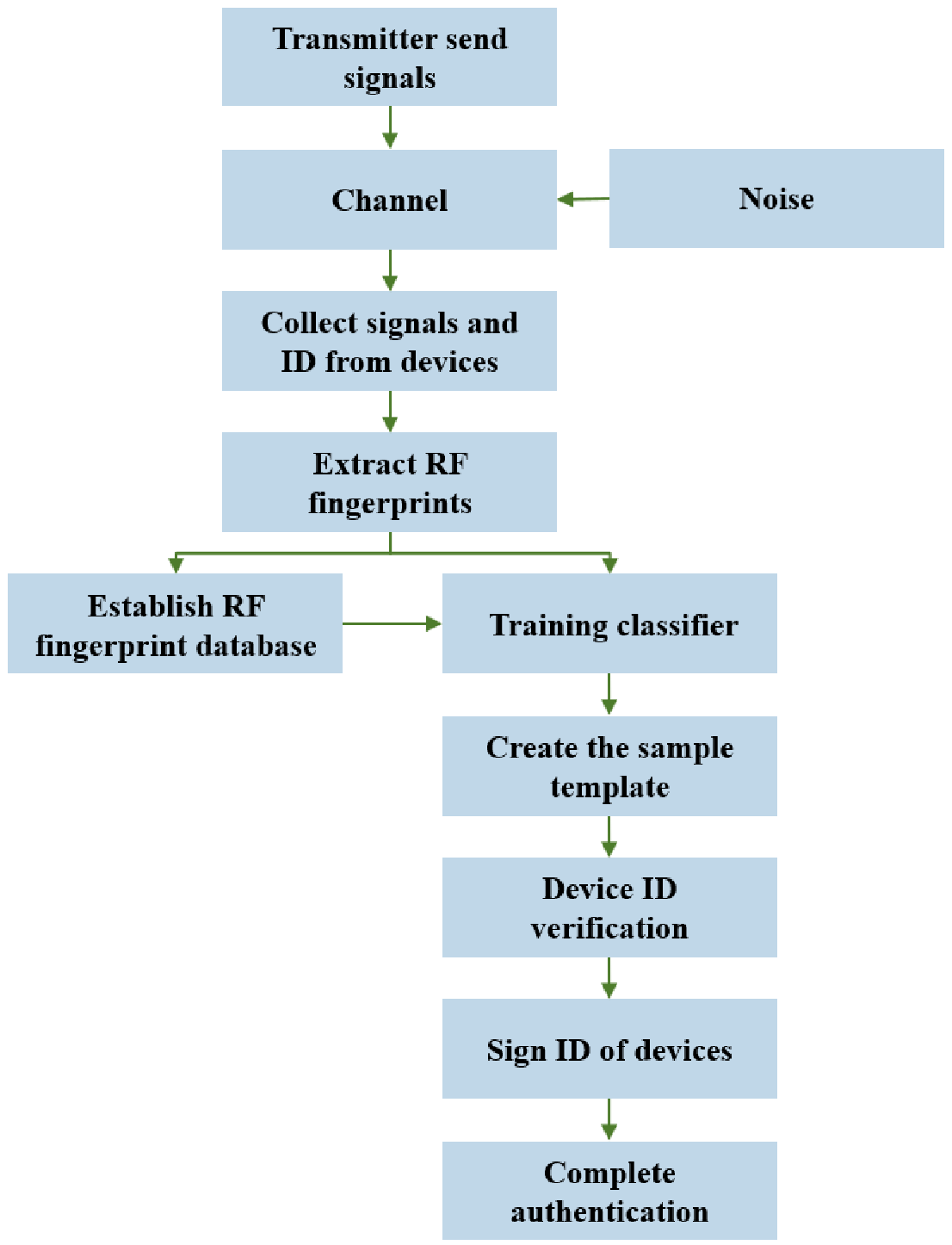}
\caption{RF fingerprint extraction step diagram.}
\label{fig2}
\end{figure}

Figure~\ref{fig2} illustrates the training process of the RF fingerprint classifier we designed. Firstly, a dataset is established with $U$ uplink RRUs and $D$ downlink RRUs. Each downlink RRU $d$ transmits random downlink data, which is received at RRU $u$ after passing through a channel with noise. The LOS path signal is removed using the known downlink data at the EDU, and tensor decomposition is applied to separate the $L$ NLOS path signals, which serve as the extracted RF fingerprints. Labels are then assigned to each NLOS path signal, corresponding to the ID of the respective downlink RRU.
Once the RF fingerprint dataset is established, 80$\%$ of the dataset is used as the training set for the classifier, while the remaining 20$\%$ is used as the testing set. The classifier is trained using the training set. During the testing phase, the testing set is used as input to the trained classifier, which outputs the ID of the downlink RRU. This establishes a mapping between the received data and the signal transmitter, allowing the EDU to determine the spatial position of the downlink RRU based on the ID.
During testing, if the ID output by the classifier matches the ID in the label, it is considered as a correct classification, otherwise, it is considered as an error. Overall, the purpose of designing this classifier is to map the separated multipath signals to the ID of the transmitting RRU, effectively determining the spatial position of the signal transmitter RRU.

In recent years, artificial intelligence methods such as deep learning and machine learning have been widely applied in the research of RF fingerprint extraction. In reference \cite{13}, the RF signal spectrum is used as input to train a neural network-based RF fingerprint recognition system using CNN. Reference \cite{15} introduces a novel hybrid network called FWSResNet, which exhibits robustness to non-smooth signals and can extract key features from noisy signals. In this study, FWSResNet achieved recognition accuracy of 93$\%$ with only 280 training samples per device, and when the number of training samples per device increased to 4200, the recognition accuracy reached 99.5$\%$. These studies strongly demonstrate the rationale and feasibility of RF fingerprint extraction in CF-RAN architecture.

To validate the effectiveness of deep learning for RF fingerprint recognition, according to reference \cite{13}, we utilized a CNN network to train the designed classifier. After training the classifier through the neural network, it quickly converged, and the judgment accuracy for RRU to match the transmitter's ID reached 92.14$\%$. Detailed training results will be provided in the simulation results section. Based on this result and the findings in reference \cite{15}, we assume that the accuracy of classifiers trained using different networks falls within the range of 90-98$\%$. It takes into account the complexity of the environment and the number of devices, with the accuracy appropriately reduced under significant environmental interference.

\subsection{Passive position estimation}
In the RF fingerprint extraction section, we separate the multipath signals using tensor decomposition algorithms \cite{gong2022multipath,lin2021tensor,zanatta2020tensor}. Assuming that the EDU utilizes RF fingerprint extraction methods, it possesses a certain accuracy in determining the signal source. The received signals after tensor decomposition can be represented as 

\begin{align}
  \label{eq7}
  \boldsymbol{y}_{d,u}^0 = \sum_{l=1}^{L} \alpha_{d, u, l}^0 \boldsymbol{e}^{-j 2 \pi {f}_{0} \tau_{d, u, l}^0 } \boldsymbol{e}^{j 2 \pi T_{x} \nu_{d, u, l}^0 } \nonumber \\
  \boldsymbol{a}\left(\varphi_{d, u, l}^0, \theta_{d, u, l}^0\right) \boldsymbol{x}_d + \mathbf{w}_{u,d}.
\end{align}
We set $D$ downlink RRUs and $U$ uplink RRUs in the scenario to perform ISAC passive sensing, taking into account the absence of interference from uplink UE signals during uplink idle periods. The $l$-th multipath signal from the $d$-th downlink RRU to the $u$-th uplink RRU is represented by $d = 1,2,...,D$ and $l = 1,2,...,L$, and the channel parameters $( {\varphi}_{d,u,l}^0,{\theta }_{d,u,l}^0,{\tau }_{d,u,l}^0,{\nu }_{d,u,l}^0 )$ can be extracted from $\boldsymbol{y}_{d,u}^0$ \cite{badiu2017variational,fleury1999channel,han2019efficient}.

The spatial position of the $d$-th downlink RRU is represented by $\boldsymbol{p}_d = [\boldsymbol{x}_d,\boldsymbol{y}_d,\boldsymbol{z}_d]^\mathbf{T}$. Similarly, the spatial position of the $u$-th uplink RRU is represented by $\boldsymbol{p}_u = [\boldsymbol{x}_u,\boldsymbol{y}_u,\boldsymbol{z}_u]^\mathbf{T}$. The position of the detected reflector is represented by $\boldsymbol{p}_r = [\boldsymbol{x}_r,\boldsymbol{y}_r,\boldsymbol{z}_r]^\mathbf{T}$, while the position of the long-standing reflector is represented by $\boldsymbol{p}_d = [\boldsymbol{x}_d,\boldsymbol{y}_d,\boldsymbol{z}_d]^\mathbf{T}$. Similarly, the spatial position of the $u$-th uplink RRU is represented by $\boldsymbol{p}_u = [\boldsymbol{x}_u,\boldsymbol{y}_u,\boldsymbol{z}_u]^\mathbf{T}$. The position of the detected reflector is represented by $\boldsymbol{p}_{r0} = [\boldsymbol{x}_{r0},\boldsymbol{y}_{r0},\boldsymbol{z}_{r0}]^\mathbf{T}$.
For the LOS path, the estimated distance from RRU $d$ to RRU $u$ is denoted as $r_{d,u,l}^0 =  v_c \tau_{d,u,l}^0$ . 
$v_c$ represents the signal transmission speed, and $r_{d,u,l} = \|\boldsymbol{p}_d - \boldsymbol{p}_u\|_2$ represents the actual distance. If the distance estimation error of the LOS path $\varepsilon_{d,u,l}^0 = |r_{d,u,l}-r_{d,u,l}^0|<\varepsilon^r$, the channel parameter estimation is considered accurate, and it can be confirmed that the signal originates from RRU $d$. Otherwise, the set of parameters is discarded. $\varepsilon^r $ is the acceptable distance error threshold.
For the NLOS paths, from the reception perspective of RRU $u$, each multipath $l$ represents a reflector. Let the estimated coordinates of the reflector be denoted as $\boldsymbol{p}_r^0 = [\boldsymbol{x}_r^0,\boldsymbol{y}_r^0,\boldsymbol{z}_r^0]^\mathbf{T}$, the estimated signal azimuth angle from the reflector as $\left(\varphi_{d, u, l}^0, \theta_{d, u, l}^0\right)$,$\varphi_{d, u, l}^0 = \arctan \frac{y_r^0-y_u}{x_r^0-x_u} $, $\theta_{d, u, l}^0=\arctan \frac{z_r^0-z_u}{r_{d,u,l}^0} $, and the actual azimuth angle as $\varphi_{d, u, l} = \arctan \frac{y_r-y_u}{x_r-x_u} $, $\theta_{d, u, l}=\arctan \frac{z_r-z_u}{r_{d,u,l}} $. The angle estimation error is denoted as $\varepsilon _{d,u}^o = \sqrt{|\varphi _{d,u,l}-\varphi _{d,u,l}^0|^2+|\theta _{d,u,l}-\theta _{d,u,l}^0|^2} $. The estimated transmission distance is $r_{d,u,l}^0=v_c \tau_{d,u,l}^0 $, and the actual transmission distance is $r_{d,u,l} = \|\boldsymbol{p}_d-\boldsymbol{p}_r\|_2+\|\boldsymbol{p}_r-\boldsymbol{p}_u\|_2$. The distance estimation error is represented by $\varepsilon_{d,u,l}^r = |r_{d,u,l}-r_{d,u,l}^0|$.
Since the coordinates of RRU $d$ and RRU $u$ are known, we can establish

\begin{align}
  \label{eq8}\
  \begin{cases}
    \left\|\boldsymbol{p}_{d}-\boldsymbol{p}_{r}^{0}\right\|_{2}+\left\|\boldsymbol{p}_{r}^{0}-\boldsymbol{p}_{u}\right\|_{2}=v_{c} \tau_{d, u, l}^{0}+\varepsilon_{u, d, l}^{r 0}\\
    \\
    \varphi_{d, u, l}^{0}=\arctan \frac{y_{r}^{0}-y_{u}}{x_{r}^{0}-x_{u}}\\
    \\
    \theta_{u, d, l}^{0}=\arctan \frac{z_{r}^{0}-z_{u}}{\left\|\boldsymbol{p}_{r}^{0}-\boldsymbol{p}_{u}\right\|_{2}}
  \end{cases}
\end{align}

Convert the equation into a form that is easier to solve

\begin{align}
  \label{eq9}\
  \begin{cases}
    \left\|\boldsymbol{p}_{d}-\boldsymbol{p}_{r}^{0}\right\|_{2}+\left\|\boldsymbol{p}_{r}^{0}-\boldsymbol{p}_{u}\right\|_{2}-v_{c} \tau_{d, u, l}^{0}=\varepsilon_{u, d, l}^{r 0} \\
    \\
    x_{r}^{0}=x_{r}^{0} \\
    \\
    y_{r}^{0}=\left(x_{r}^{0}-x_{u}\right) \tan \varphi_{d, u, l}^{0}+y_{u} \\
    \\
    z_{r}^{0}=\tan \theta_{u, d, l}^{0} \sqrt{\left(x_{u}-x_{r}^{0}\right)^{2}+\left(y_{u}-y_{r}^{0}\right)^{2}}+z_{u}
  \end{cases}
\end{align}

When $\varepsilon_{u, d, l}^{r 0}={\varepsilon_{min}}_{u, d, l}^{r 0}$, indicating that the estimated position of the reflector is $\boldsymbol{p}_r^0 = [\boldsymbol{x}_r^0,\boldsymbol{y}_r^0,\boldsymbol{z}_r^0]^\mathbf{T}$, if a reflector is detected by multiple uplink RRUs due to their close proximity, the EDU will aggregate multiple detection results into a position estimation vector $\boldsymbol{p}_r^0$ for that reflector. The average of the $P$ detection results $\boldsymbol{p}_r^0=[\boldsymbol{p}_{r,1}^0,\boldsymbol{p}_{r,2}^0,...,\boldsymbol{p}_{r,P}^0]$ for that reflector is $\bar{\boldsymbol{p}}_{r}^0=\frac{\sum_{p = 1}^{P}\boldsymbol{p}_{r,p}^0}{P} $ as the final detection result, defined as the average perception error $\varepsilon _p = \|\bar{\boldsymbol{p}}_{r}^0-\boldsymbol{p}_{r}\|_2$. If the reflector is far away, it is considered as a different reflector.

Most reflectors remain fixed in position for a long time, such as buildings, plants, the ground, and parked vehicles. Therefore, before separating the signals, the influence of irrelevant targets $\hat{Y}_{NLOS}$ can be eliminated using historical data. This allows the perception system to focus its attention more effectively on moving and changing targets, which are often the objects that the perception system needs to serve. At the same time, it significantly reduces the computational complexity of NLOS channel parameter estimation.

\section{Simulation results}
We consider a CF-RAN scenario with $3000~\mathrm{m} \times 3000~\mathrm{m} \times 60~\mathrm{m}$ and perform 
a Monte Carlo simulation with $Q$ RRUs and a $K$ reflectors, including $K_{a}$ long-standing 
interfering reflectors and $K_{b}$ short-term target reflectors, where the $Q$ RRUs are assigned 
different uplink and downlink tasks at different times through CPU control, and the downlink 
data information sent at different times is randomly generated. In the same time slot, the RRU 
assigned to the downlink task is responsible for transmitting signals, while the RRU assigned 
to the uplink task is responsible for receiving signals and resolving the sensed signals. 
These RRUs are controlled by the EDU and can know the information that can be shared among 
them through the EDU, such as the location and the content of the signals sent. 

\begin{figure}[!ht]
  \centering
  \includegraphics[width=1\linewidth]{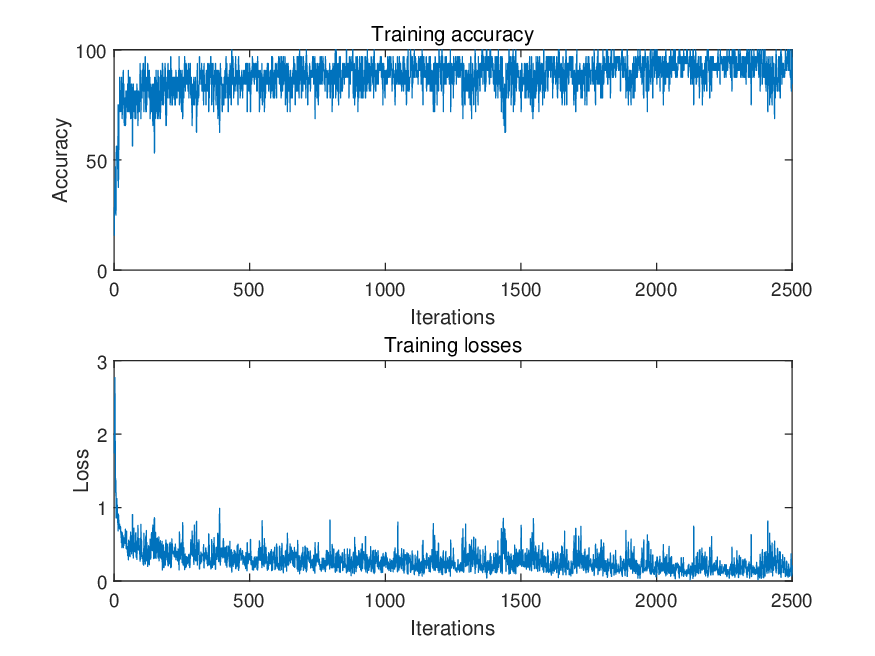}
  \caption{CNN network training convergence graph. Number of samples 8000, symbols 256, large noise.}
  \label{fig3}
  \end{figure}

\begin{figure}[!ht]
  \centering
  \includegraphics[width=1\linewidth]{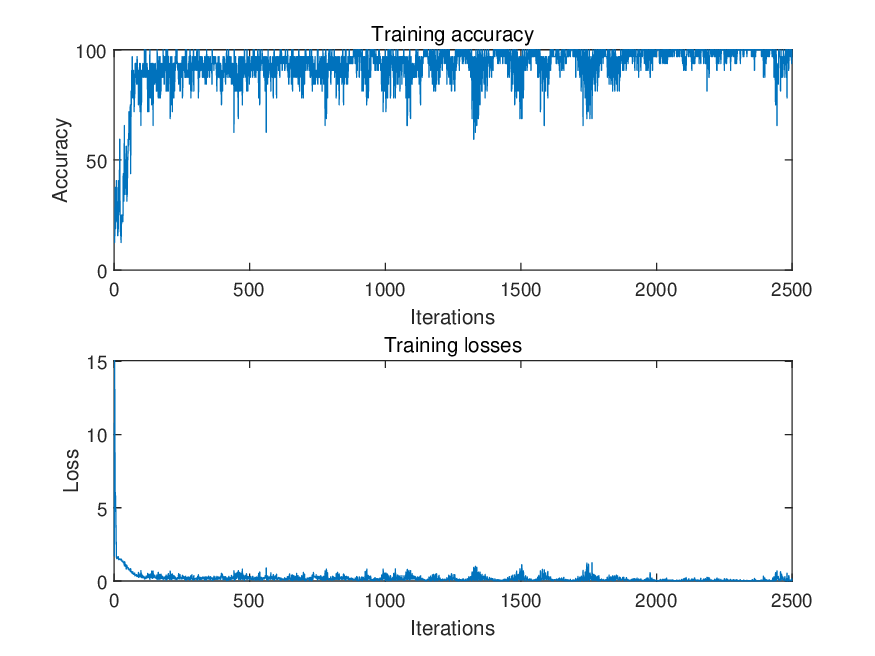}
  \caption{CNN network training convergence graph. Number of samples 8000, symbols 1000, small noise.}
  \label{fig4}
\end{figure}

As shown in Figure~\ref{fig3}, we conducted neural network training of the designed classifier using a CNN network. In a scenario with Q=5 and D=5, the uplink RRUs received NLOS signals reflected by reflectors. After eliminating the known transmitted data, the data was used for training. In the simulation, when the power spectral density of the RF fingerprint signal generated by hardware impairments was 0.3 times the channel noise, with 8000 training samples per downlink RRU and 256 signal symbols, the accuracy of the receiving RRU in determining the source of the tensor decomposition reached 81.54$\%$. When the power spectral density of the RF fingerprint signal was 0.4 times the channel noise, as Figure~\ref{fig4}, with 8000 training samples and 1000 signal symbols, the accuracy of the receiving RRU in determining the source reached 92.14$\%$. Lower channel noise led to better training results, and the channel noise depended on the accuracy of channel estimation.

The above simulations support our hypothesis that after training, the RRU can have a list of RF fingerprints for each other and can accurately determine the source of the signal. In the simulation of RRU perception accuracy, we introduced zero-mean Gaussian distribution errors with different standard deviations into the multipath parameter information obtained from RRU separation. We simulated the performance of the passive positioning estimation algorithm under the real decoding algorithm and channel environment.

\begin{figure}[!ht]
\centering
\includegraphics[width=1\linewidth]{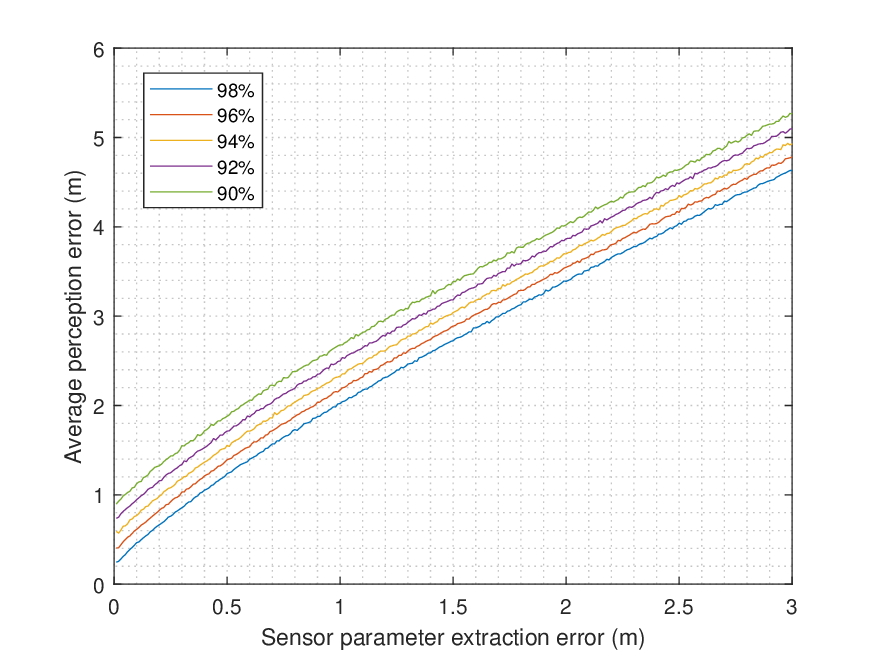}
\caption{Relationship between estimation error of perceptual parameters and localization error of reflector with different RF fingerprint extraction accuracy (98$\%$-90$\%$).}
\label{fig5}
\end{figure}
According to the description in Section 3.3, we define the parameter estimation errors as $\varepsilon _{u,d,l}=[\varepsilon _{u,d,l}^r,\varepsilon _{u,d,l}^o]$, representing the distance estimation error and the direction angle estimation error. The average perception error is defined as $\varepsilon _{p}=\| \bar{\boldsymbol{p}}_r^0-\boldsymbol{p}_r\|_2$.

As in Figure~\ref{fig5}, we simulated the reflector localization error under the influence of different RF 
fingerprint judgment accuracy and parameter estimation error of the signal at the receiver 
side. We ignored those obviously wrong results in the RRU localization phase, such as the passive 
objects with high signal reception power but at a relative distance of more than $300~\mathrm{m}$, which is obviously 
a work that can be done in the channel estimation. Simulation results show that the accuracy 
of passive sensing depends more on the accuracy of channel parameter estimation. When the 
parameter estimation Cartesian coordinate system error is within $0.5~\mathrm{m}$, the total accuracy 
error can be kept within $2~\mathrm{m}$ despite the RF fingerprint extraction accuracy of only 90$\%$, 
and the total accuracy error can even be within 1m when the RF fingerprint information 
accuracy is increased.

\begin{figure}[!ht]
  \centering
  \includegraphics[width=1\linewidth]{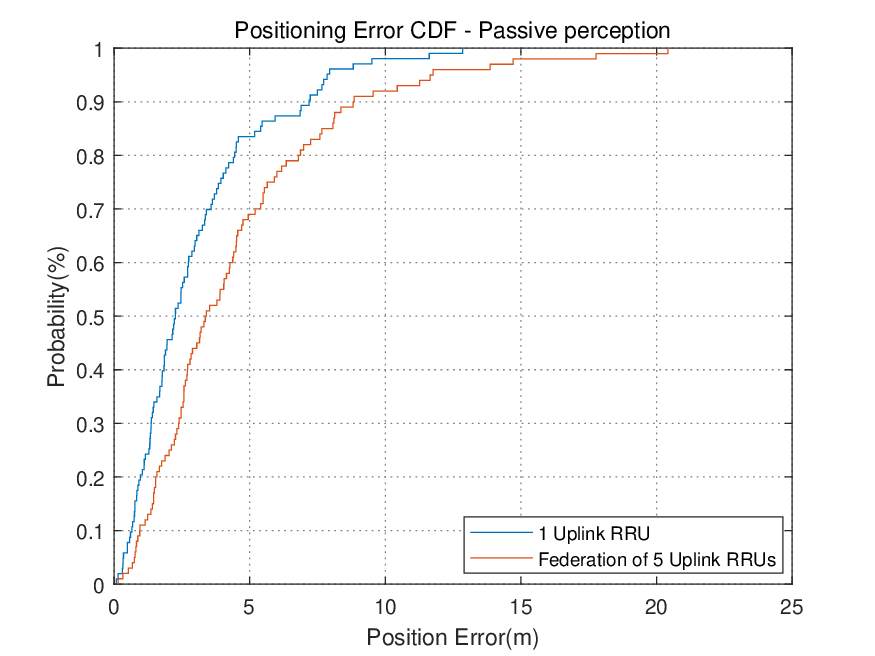}
  \caption{The cumulative distribution function (CDF) plot of the position estimation error shows that the performance of the multi-RRU simultaneous co-frequency joint sensing is slightly inferior to that of the single RRU sensing, but it is relatively close in terms of performance.}
  \label{fig6}
  \end{figure}

As shown in Figure~\ref{fig6}, since there is no existing work on passive sensing using the same ISAC signal with multiple RRU in the CF-RAN architecture, we compared the perception performance of our proposed architecture with the case where a single downlink RRU sends ISAC signal and multiple uplink RRU receive the ISAC signal reflected by the reflectors. Simulation results demonstrate that the architecture with a single downlink RRU sending ISAC signal and multiple uplink RRU receiving the ISAC signal reflected by the reflectors indeed has advantages in terms of passive sensing accuracy. However, the multi-RRU approach utilizing RF fingerprint identification and classification can also achieve comparable perception accuracy. Nevertheless, under the ISAC operational mode of this architecture, the communication, data rate, and signal-to-noise ratio are evidently not as good as the case with multiple downlink RRU working in a coordinated manner. This is the inherent advantage of a distributed communication architecture.

\section{Conclusion}

This paper proposes an innovative design concept that integrates mobile communication and passive sensing in the Cell-Free RAN architecture. It achieves efficient passive sensing based on RF hardware without occupying communication bandwidth or compromising communication performance. Additionally, it addresses the problem of source identification in multi-RRU sensing in the CF-RAN architecture by discussing an RRU source identification method based on RF fingerprint extraction. Furthermore, a computational method for passive estimation of RRU positions is proposed, which effectively addresses the problem of position estimation for reflections with low computational complexity. Simulation results validate the passive sensing performance in scenarios with multiple RRUs and multiple reflections. The paper demonstrates the effectiveness of a CNN-based RF fingerprint classification model for source identification in multi-RRU collaborative passive sensing. However, there are limitations in this work, such as the need for more effective parameter extraction from received multipath signals and more efficient AI-based RF fingerprint extraction algorithms. These limitations will be addressed in future research efforts.

\section*{ACKNOWLEDGEMENT}
\label{ACKNOWLEDGEMENT}
This work was supported in part by the National Key Research and Development Program under Grant (2021YFB2900300), and by the National Natural Science Foundation of China (NSFC) under Grants 61971127, 61871122, by the Southeast University-China Mobile Research Institute Joint Innovation Center, and by the Major Key Project of PCL (PCL2021A01-2).

\bibliographystyle{gbt7714-numerical}
\bibliography{myref}

\biographies

\begin{CCJNLbiography}{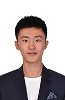}{Jingxuan Yu}
received the B.S degree in Information Engineer from Southeast University in 2021 and is currently pursuing his M.S degree in the School of Information Science and Engineering at Southeast University. His main research interests are in communication and sensing integration. His supervisor is Professor Li Jiamin.
\end{CCJNLbiography}

\begin{CCJNLbiography}{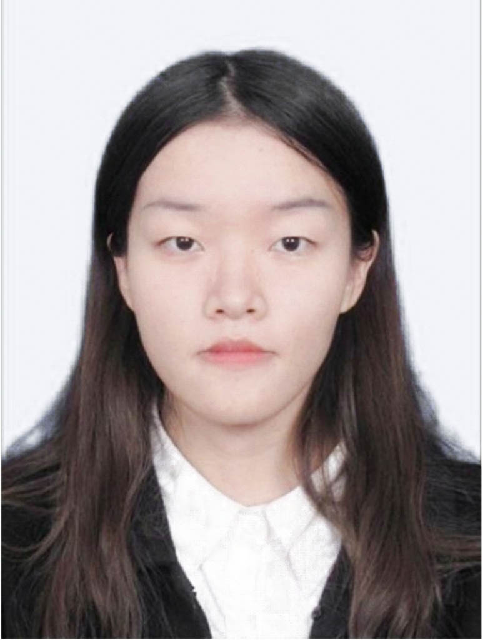}{Fan Zeng}
received the B.S. degree incommunication engineering from Shang-hai University, Shanghai, China, in 2021.She is currently pursuing the M.S. de-gree with the College of Information Sci-ence and Engineering,Southeast Univer-sity. Her research interests include inte-grated sensing and communication.
\end{CCJNLbiography}

\begin{CCJNLbiography}{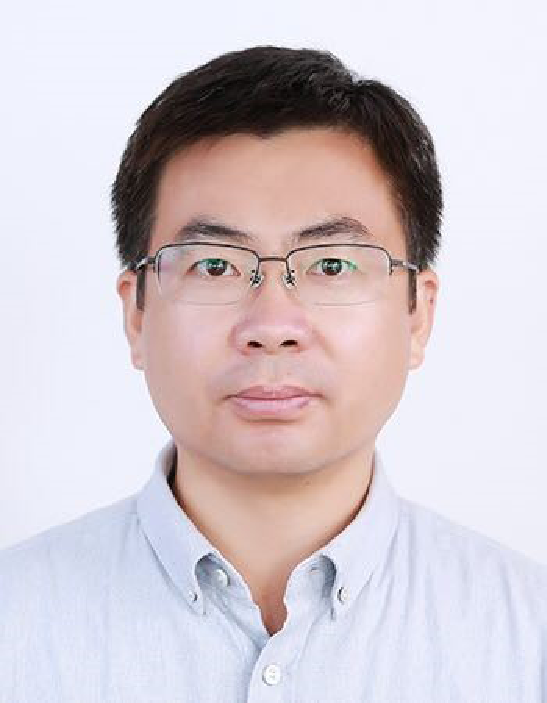}{Jiaming Li}
received the B.S. and M.S. degrees in communication and information systems from Hohai University, Nanjing, China, in 2006 and 2009, respectively, and the Ph.D. degree in information and communication engineering from Southeast University, Nanjing, China, in 2014. He joined the National Mobile Communications Research Laboratory, Southeast University, in 2014, where he has been an Associate Professor since 2019. His research interests include massive MIMO, distributed antenna systems, and multi-objective optimization.
\end{CCJNLbiography}

\begin{CCJNLbiography}{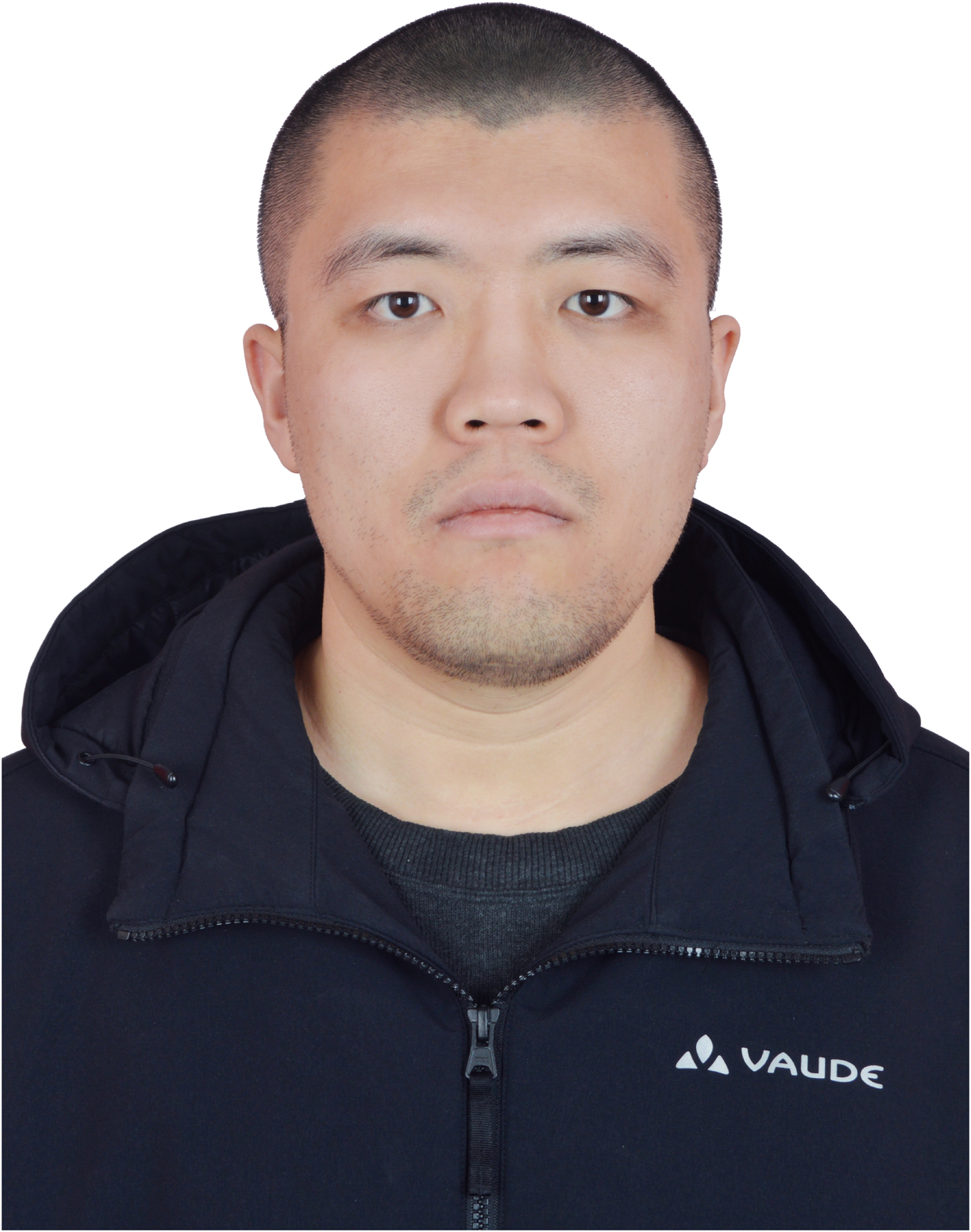}{Feiyang Liu}
received the B.S. and M.S. degrees from Xidian University, Xi'an, China, in 2014 and 2017, respectively. He joined in the 54th Research Institute of China Electronics Technology Group Corporation, in 2017.His research interests include UAV broadband data chain.
\end{CCJNLbiography}

\begin{CCJNLbiography}{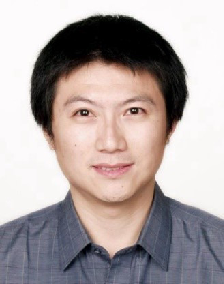}{Pengcheng Zhu}
received the B.S and M.S. degrees in electrical engineering from Shandong University, Jinan, China, in 2001 and 2004, respectively, and the Ph.D. degree in communication and information science from the Southeast University, Nanjing, China, in 2009. He has been a lecturer with the national mobile communications research laboratory, Southeast University, China, since 2009. His research interests lie in the areas of communication and signal processing, including limited feedback techniques, and distributed antenna systems.
\end{CCJNLbiography}

\begin{CCJNLbiography}{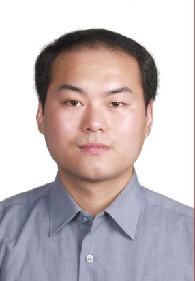}{Dongming Wang}
received the B.S. degree from Chongqing University of Posts and Telecommunications, Chongqing, China, the M.S. degree from Nanjing University of Posts and Telecommunications, Nanjing, China, and the Ph.D. degree from the Southeast University, Nanjing, China, in 1999, 2002, and 2006, respectively. He joined the National Mobile Communications Research Laboratory, Southeast University, in 2006, where he has been an Associate Professor since 2010. His research interests include turbo detection, channel estimation, distributed antenna systems, and large-scale MIMO systems.
\end{CCJNLbiography}
    
\begin{CCJNLbiography}{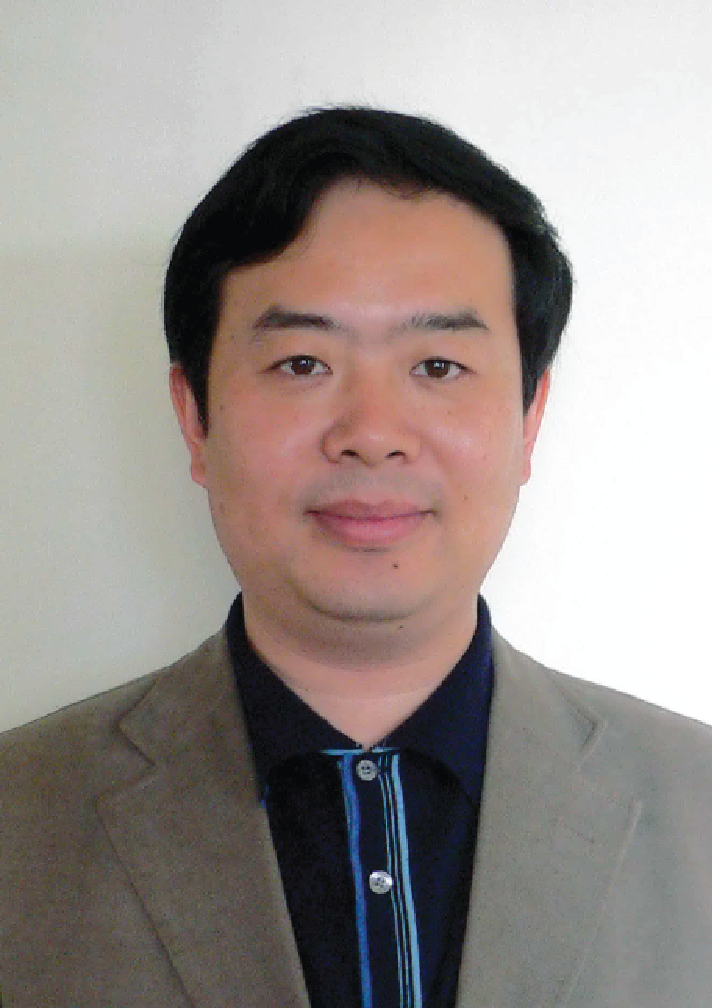}{Xiaohu You}
(Fellow, IEEE) received the B.S., M.S. and Ph.D. degrees in electrical engineering from Nanjing Institute of Technology, Nanjing, China, in 1982, 1985, and 1989, respectively. From 1987 to 1989, he was with Nanjing Institute of Technology as a Lecturer. From 1990 to the present time, he has been with Southeast University, first as an Associate Professor and later as a Professor. His research interests include mobile communications, adaptive signal processing, and artificial neural networks with applications to communications and biomedical engineering. He is the Chief of the Technical Group of China 3G/B3G Mobile Communication R \& D Project. He received the excellent paper prize from the China Institute of Communications in 1987 and the Elite Outstanding Young Teacher Awards from Southeast University in 1990, 1991, and 1993. He was also a recipient of the 1989 Young Teacher Award of Fok Ying Tung Education Foundation, State Education Commission of China.
\end{CCJNLbiography}

\end{document}